\documentstyle[12pt,eq]{article}

\title{Statistical Properties of the Zeros of Zeta Functions -- Beyond the
Riemann case}
\author{}
\date{E. Bogomolny and P. Leb{\oe}uf\\
\vspace{0.1in}
Division de Physique Th\'eorique\footnote{Unit\'e  de Recherche
des Universit\'es de Paris XI et Paris VI associ\'ee au CNRS}\\ Institut de
Physique Nucl\'eaire\\91406 Orsay Cedex, France \vspace{0.15in} \\}

\newcommand{\beq}{\begin{equation}}
\newcommand{\eeq}{\end{equation}}
\newcommand{\beeqn}{\begin{eqnarray}}
\newcommand{\eeqn}{\end{eqnarray}}
\newcommand{\rs}{\zeta (s)}
\newcommand{\ds}{L (s, \chi) }
\newcommand{\dt}{d (t, \chi) }
\newcommand{\davc}{d_{\rm av} (t, \chi) }
\newcommand{\dav}{d_{\rm av} }
\newcommand{\ddav}{d^2_{\rm av} (t, \chi) }
\newcommand{\dosc}{ d_{\rm osc} (t, \chi) }
\newcommand{\zs}{Z (t, \chi)}
\newcommand{\rd}{R_2 (\epsilon)}
\newcommand{\rdo}{R_2^{\rm osc} (\epsilon)}
\newcommand{\wc}{W_\chi}
\newcommand{\ts}{\Theta_\chi (t)}
\newcommand{\ah}{ \alpha (h)}
\newcommand{\cn}{ \chi (n)}
\newcommand{\cq}{ \chi (q)}
\newcommand{\cp}{ \chi (p)}
\newcommand{\cbpp}{ {\bar \chi} (p')}
\newcommand{\sn}{ \sum_{n=1}^{\infty}}
\newcommand{\sq}{ \sum_{q=1}^{k-1}}
\newcommand{\cl}{${\cal L}_c \,$}

\newcommand{\eqinf}{\stackrel{t\rightarrow\infty}{\simeq}}
\renewcommand{\theequation}{\arabic{section}.\arabic{equation}}

\begin{document}
\baselineskip 0.3in
\setcounter{page}{0}
\maketitle

\vspace{0.5in}

{\Large{\bf Abstract:}}
{We investigate the statistical distribution of the zeros of Dirichlet
$L$--functions both analytically and numerically. Using the Hardy--Littlewood
conjecture about the distribution of primes we show that the two--point
correlation function of these zeros coincides with that for eigenvalues of
the Gaussian unitary ensemble of random matrices, and that the distributions of
zeros of different $L$--functions are statistically independent.
Applications of these results to Epstein's zeta functions are
shortly discussed.}

\vspace{0.9in}

\noindent PA numbers: 0545; 0365; 0250\\

\noindent Short Title: Statistical Properties of the Zeros of Zeta Functions

\pagebreak

\def\sfrac#1#2{{\textstyle{#1 \over #2}}}
\def\ie{{\it i.e.}}
\def\eg{{\it e.g.}}
\def\Eg{{\it E.g.}}
\def\etal{{\it et al.}}
\def\gtwid{\,{\raise.3ex\hbox{$>$\kern-.75em\lower1ex\hbox{$\sim$}}}\,}
\def\ltwid{\,{\raise.3ex\hbox{$<$\kern-.75em\lower1ex\hbox{$\sim$}}}\,}
\def\half{\frac{1}{2}}
\def\third{\frac{1}{3}}
\def\fourth{\frac{1}{4}}
\def\eighth{\frac{1}{8}}
\def\mi{{\rm i}}
\def\me{{\rm e}}
\def\[{\left [}
\def\]{\right ]}
\def\({\left (}
\def\){\right )}
\def\lb{\left |}
\def\rb{\right |}
\def\cbl{\left\{}
\def\cbr{\right\}}
\def\sss#1{{\scriptscriptstyle #1}}
\def\ss#1{{\scriptstyle #1}}
\def\ssr#1{{\sss{\rm #1}}}
\def\undertext#1{$\underline{\hbox{#1}}$}
\def\intl{\int\limits}
\def\Tr{\mathop{\rm Tr}}
\def\ket#1{{\,|\,#1\,\rangle\,}}
\def\bra#1{{\,\langle\,#1\,|\,}}
\def\braket#1#2{{\,\langle\,#1\,|\,#2\,\rangle\,}}
\def\expect#1#2#3{{\,\langle\,#1\,|\,#2\,|\,#3\,\rangle\,}}
\def\eq#1#2#3{Eq.(\ref #1#2#3)}
\def\re#1#2#3{(\ref #1#2#3)}
\def\eps{\epsilon}

\section{Introduction}
\setcounter{equation}{0}

Statistical reasoning and the modelization of physical phenomena by random
processes have taken a major place in modern physics and mathematics. A
physical example of this is the emergence of random behaviour from purely
deterministic laws, as in classically chaotic Hamiltonian systems (see
e.g.~\cite{ll}). Randomness also enters in the quantum version of these systems
(see e.g.~\cite{lh}). In fact, the statistical properties of the semiclassical
spectrum of fully chaotic systems are, in the universal regime, in good
agreement with those obtained from an ensemble of random matrices
\cite{mehta,bohigas}. The Gaussian orthogonal ensemble (GOE) statistics applies
to systems which are chaotic and have (generalized) time--reversal symmetry,
while the Gaussian unitary ensemble (GUE) statistics are appropriated to
describe systems which are chaotic and without time--reversal invariance. In
particular, the GUE two--point correlation function is (after normalization of
the average spacing between eigenvalues to unity)
\beq \label{11}
R_2^{GUE} (\eps) = 1 - \frac{\sin^2 (\pi \eps)}{\pi^2 \eps^2},
\eeq
or its Fourier transform, the two--point form factor, has the form:
\beq \label{12}
K_2^{GUE} (\tau) = \left\{ \begin{array}{ll}
         |\tau| & \mbox{if $|\tau|<1$} \\
           1    & \mbox{if $|\tau|>1$.} \end{array} \right.
\eeq

A particularly interesting example of applying statistical considerations
to a pure mathematical object is provided by the Riemann zeta function.
This function is defined by a series
\beq \label{13}
\rs = \sn \frac{1}{n^s}
\eeq
which converges for $Re(s)>1$ and can be analytically continued to the
whole complex plane \cite{tit}. The region $0<Re(s)<1$ is
called the critical strip and it was proved~\cite{reich} that in this
region the Riemann zeta function has properties of a random function.

The Riemann hypothesis asserts that all the complex zeros of $\rs$ lie on the
critical line $Re(s)=1/2$, which we henceforth denote by \cl. Assuming the
Riemann hypothesis, Montgomery \cite{montgomery,gm} concluded that
asymptotically (i.e., for large values of $Im(s)$), the form factor of the
critical Riemann zeros coincides, for $|\tau|<1$, with the GUE result
(\ref{12}) (and conjectured that the agreement holds for arbitrary $\tau$).
More recently, some spectacular numerical results by Odlyzko \cite{odlyzko}
strongly support that conjecture. Assuming certain number theoretical
hypotheses on the correlations between prime numbers to hold, in Ref.\cite{gm}
and later on in Ref.\cite{keating} it was shown that the main term of the
two--point correlation function for the critical zeros of the Riemann zeta
function does coincide with (\ref{12}). In the context of "quantum chaos", an
analogue of Montgomery's result was found by Berry \cite{berry}, who also
discussed the validity of the random matrix theory.

These two apparently disconnected physical and mathematical results have a
common root in a formal analogy between the density of Riemann zeros expressed
in terms of prime numbers (cf. Eqs.(\ref{213}) below) and an asymptotic
approximation of the quantum spectral density in terms of classical periodic
orbits (the Gutzwiller trace formula~\cite{gutz}). This analogy has been
fruitful for both mathematical and physical fields. For example, the
correlations between prime numbers (the Hardy--Littlewood conjecture) inspired
some work on the correlations between periodic orbits \cite{ks}. In the
opposite direction, the statistical non--universalities of the spectral
density, related to short periodic orbits, were successfully transposed to the
Riemann zeta function \cite{berry2}.

It is thus clear that zeta functions are good models for investigating level
statistics and the semiclassical trace formula. There are many generalizations
of the Riemann zeta function \cite{encyclopedia}, and since very little is
known about their zeros it is of interest to investigate their distribution. In
\cite{ozluk} the analogue of Montgomery's result was proved for the average of
all Dirichlet $L$--functions having the same (large) modulus, while $Im(s)$ is
kept constant. The purpose of this paper is to study, in the limit of large
$Im(s)$, the statistical properties of zeros of {\sl individual} Dirichlet
$L$--functions having arbitrary modulus and character.

After a brief introduction (Section 2), in Section 3 we prove that the main
asymptotic term of the two--point correlation function of the non--trivial
zeros of an arbitrary Dirichlet $L$--function agrees with GUE. We also prove
the statistical independence of the zeros of $L$--functions having different
character and/or different modulus, i.e.  the zeros of their product behave
like the superposition of uncorrelated GUE--sets. The demonstrations assume
that the Hardy-Littlewood conjecture on the correlations between prime numbers
is valid. The application of these results to the distribution of the zeros of
the zeta function of positive binary quadratic forms, a particular case of the
Epstein zeta function, is shortly discussed in Section 4.

\section{Dirichlet $L$--functions}
\setcounter{equation}{0}

Dirichlet $L$--functions are natural generalizations  of the Riemann zeta
function (\ref{13}). When $Re(s)>1$ they are defined by a series (see, e.g.
\cite{bateman})
\beq \label{23}
\ds = \sn \frac{\cn}{n^s} = \prod_p \( 1-\frac{\chi (p)}{p^{s}}\)^{-1}.
\eeq
where the product is taken over all primes $p$.

Given an arbitrary integer $k$ (called the modulus), a function $\cn$ (called a
Dirichlet character mod $k$) is a complex function of positive integers
satisfying:
(i) $\chi (nm) = \cn \, \chi (m)$,
(ii) $\chi (n)=\chi (m)$ if $n\equiv m$ mod $k$,
(iii) $\chi (n)=0$ if $(n,k)\neq 1$,
where $(n,k)$ denotes the highest common divisor of $n$ and $k$.

A character is called {\sl principal} and denoted by $\chi_0$ if $\chi_0 = 1$
when $(n,k)=1$ and $\chi_0=0$ otherwise; the corresponding $L$--function
essentially reproduces the Riemann zeta function. In fact
$$
 L ( s, \chi_0)=\rs  \prod_{p|k} \(1-p^{-s}\),
$$
where the product is taken over all prime factors of $k$. It also follows from
the above definitions that $\chi(1)=1$ and $\[\chi(k-1)\]^2=\[\chi(-1)\]^2=1$.

In general $k$ can be any integer number. The total number of different
characters modulo $k$  is given by Euler's function $\phi(k)$
(the number of positive integers prime to, and not exceeding $k$). The value
$\cn$ is different from zero iff $(n,k)=1$ and its $\phi(k)-th$ power equals
one. Table 1 provides a list of non--principal characters for $k=4$ and $5$, to
be later on used in the numerical computations. (Detailed tables of characters
can be found in \cite{tc}).

A character $\chi$ mod $k$ is called nonprimitive if there is a divisor $k'$ of
$k$ such that when $n'\equiv n$ mod $k'$, $\chi (n')=\chi (n)$. Otherwise the
character is called primitive. For primitive characters Dirichlet
$L$--functions satisfy the functional equation \cite{dh,davies}:
\beq \label{26}
\xi (s,\chi) = (- \mi)^a \, \wc \; \xi (1-s, {\bar \chi})
\eeq
where
$$
\xi (s,\chi) = \( \frac{k}{\pi} \)^{s/2} \Gamma\(\frac{s+a}{2}\)\, \ds
$$
and $a=\[1-\chi (-1) \]/2$. $\wc$ is a complex number of unit modulus which,
for a given character, is a constant
\beq \label{27}
\wc = \frac{1}{\sqrt{k}} \sq \me^{2 \pi \mi q/k} \, \cq.
\eeq

Like in the Riemann case, the functional equation allows to define a real
function on the critical line \cl (where according to the generalized Riemann
hypothesis should lie all non--trivial zeros of $L$--functions):
\beq
\zs = \me^{- \mi \ts /2} \, L ( 1/2 - \mi t, \chi) \label{29}
    = \left. \sum_{n=1}^{\infty}\right.^\prime \frac{1}{\sqrt{n}}
      \cos \[ t \ln n - \ts/2 + \arg \cn\] \label{210}
\eeq
where the symbol ${\sum}^{'}$ indicates that the summation is done over all
terms for which $\chi (n)\neq 0$ and
$$
\ts = \arg \wc + t \ln \( \frac{k}{\pi} \) - 2 \arg \[\Gamma
\(\frac{1+2a}{4} -\frac {\mi}{2} t \)\]-\frac{a\pi}{2}. $$
Asymptotically
\beq \label{211}
\ts \eqinf \arg \wc + t\[\ln\(\frac{kt}{2\pi}\)
-1\]-\frac{\pi}{4} + {\cal O}(t^{-1}).
\eeq
An approximate functional equation (the analogue of the Riemann--Siegel formula
for $\rs$) also holds for $L$--functions in the large--$t$
limit~\cite{siegel,davies}. It takes the form of a resummation of the series:
instead of the infinite sum (\ref{210}),  $\zs$ can be written as a truncated
sum
\beq \label{212}
\zs \simeq 2 \left.\sum_{n=1}^N\right.^\prime \frac{1}{\sqrt{n}} \cos \[ t
\ln n - \ts/2 + \arg \cn\] + {\cal O}(t^{-1/4}),
\eeq
where $N=\[\sqrt{\frac{t}{2\pi k}}+\half\]k$ (the square brakets denote here
integer part). An explicit form for the correction terms in Eq.(\ref{212}) can
be found in \cite{siegel,davies}. This expression is particularly useful for
numerical computations since we need to find the real zeros of a real function
expressed as a finite sum of oscillating terms, their number being proportional
to $\sqrt{t}$.

As usual (see for example \cite{keating}), one  can express the density of
zeros lying on \cl as a sum of an average part and a fluctuating part, $\dt =
\davc + \dosc$. The result is
\begin{subeqnarray} \label{213}
&& \davc = \frac{1}{2\pi} \frac{d \ts}{d t} \eqinf \frac{1}{2\pi} \ln \(kt/2
\pi
\) \\ \label{214}
&& \dosc = -\frac{1}{\pi} {\sum_p}^{'} \sum_{m=1}^\infty \frac{\ln p}
{\me^{\frac{m}{2}\ln p}} \cos \[ m \(t \ln p + \arg \cp \)\].
\end{subeqnarray}

This decomposition is analogous to the Gutzwiller trace formula \cite{gutz},
where the quantum spectral density is expressed as a sum of an average part and
a fluctuating part. The average part of the level density is given by the
so--called Weyl or Thomas--Fermi term and is related in the lowest order
approximation to the derivative with respect to the energy of the volume of the
classical phase space energy shell. The oscillating term is expressed as a sum
over all the periodic orbits. Formally, Eq.(2.7b) is analogous to this latter
term for a system whose periodic orbits are all isolated and unstable
\cite{gutz}. In this analogy, the prime numbers are identified with periodic
orbits (whose lengths are given by the logarithm of the prime numbers) and the
characters, being pure phase factors for $(p,k)=1$, may be interpreted as
Maslov indices, a quantity classically related to the focusing of a flux tube
surrounding the periodic orbit. In part 5, however, we provide
a different (and perhaps more relevant) interpretation of the characters
in Eq.(2.7b) in terms of symmetries.

\section{Two--point correlation function}
\setcounter{equation}{0}

Let us formally insert the density of zeros given by Eqs.(\ref{213}) into the
definition of the two--point correlation function
\beq \label{31}
\rd = \langle d (t,\chi) \, d (t+\eps,\chi) \rangle
\eeq
and express $\rd$ as a sum over prime numbers. In (\ref{31}) the bracket
$\langle \; \rangle$ denotes an average over an interval $\Delta t$ which is
unavoidable when discussing statistical properties of a given function. Our
purpose is to compute $\rd$ in the large--$t$ limit and, accordingly, we choose
the smoothing interval such that $1\ll \Delta t \ll t$ \cite{berry3}. Thus,
any oscillating term with period smaller than ${\cal O}(1)$ will be washed away
by the averaging procedure. Then, from (\ref{31}) and (\ref{213})
\beq \label{32}
\rd \approx \ddav + \rdo
\eeq
where
\beeqn
\rdo &=& \langle \dosc \, d_{\rm osc} (t+\eps, \chi) \rangle \nonumber \\
&=& \frac{1}{4 \pi^2} \langle \sum_{p,p'} \sum_{m,m'=1}^{\infty}
\frac{\cp \cbpp \ln p \ln p'}{p^{m/2} \(p'\)^{m'/2}}
\exp\[\mi t \(m\ln p - m' \ln p' \) - \mi m \eps \ln p'\] \rangle
+ {\rm c.c.}. \nonumber
\eeqn
We must now show that $\rdo $ reproduces the second term in the r.h.s.
of Eq.(\ref{11}). The main lines of our proof follow those developed in
\cite{keating} for the Riemann zeta function.

The average in $\rdo$ is not zero since the difference $m \ln p - m' \ln p'$
can be arbitrarily small (for large values of $p$ and $p'$) and can produce
oscillating terms whose period is of order $\Delta t$ or bigger. Moreover, the
sums are convergent for values of $m, m'$ bigger than one. So we restrict to
$m=m'=1$
\beq \label{33}
\rdo \approx \frac{1}{4 \pi^2} \langle \sum_{p,p'}
\frac{\cp \cbpp \ln p \ln p'}{p^{1/2} \(p'\)^{1/2}}
\exp\[\mi t \(\ln p - \ln p' \) - \mi \eps \ln p'\] \rangle + {\rm c.c.}.
\eeq
Now we split the double sum into two parts $\sum_{p, p'} = \sum_{p=p'} +
\sum_{p
\neq p'}$, and first compute the diagonal part. Since $|\chi (p)|=1$ if $(p,k)
=1$, then
$$
\rdo_{\rm diag} = \frac{1}{4 \pi^2} {\sum_p}^{'} \( \frac{\ln^2 p}{p}
\me^{-\mi \eps \ln p}\) + {\rm c.c.}.
$$

Taking into account that only for a finite number of primes $(p,k)\neq 1$, it
follows that one can replace the sum over primes for which $\chi (p)\neq 0$ by
an integral, using the usual prime number theorem for the density of primes
(see e.g.~\cite{tit}). Putting $\tau = (\ln p) / 2 \pi$ one obtains:
\beq \label{34}
\rdo_{\rm diag} \approx \int_0^\infty d \tau \, \tau \, \me^{-2 \pi \mi \eps
\tau}
+{\rm c.c.} = - \frac{1}{2 \pi^2 \eps^2}.
\eeq
This contribution reproduces the non--fluctuating part of $R_2^{GUE}$.
Alternatively, the contribution of (\ref{34}) to the form factor is $\tau$, in
agreement with Eq.(\ref{12}) for $|\tau|<1$.

In the off--diagonal part of (\ref{33}) the smoothing will wash away terms
for which $\Delta t \ln (p/p') \geq 1$, and $\Delta t \rightarrow \infty$ as
$t \rightarrow \infty$. The main contribution then comes from large values
of $p$ and $p \sim p'$. Putting $p' = p + h$ and assuming that $h \ll p$
the off--diagonal part reads
\beq \label{35}
\rdo_{\rm off} \approx \frac{1}{4 \pi^2} \sum_{p} \frac{\ln^2 p}{p} \me^{-\mi
\eps \ln p}
\langle \sum_h \cp {\bar \chi} (p+h) \me^{-\mi t h/p} \rangle + {\rm c.c.},
\eeq
where the external sum is taken over all primes and the internal one is taken
over integers $h$ such that $p+h$ is a prime. To proceed further we need some
information about the pair correlation function between prime numbers. An
important conjecture related to those correlations is due to Hardy and
Littlewood \cite{hl,lc}, which expresses the density $\Lambda_h (X)$ of primes
$p$ lying between $X$ and $X + dX$ such that $p+h$ is also a prime.

According to that conjecture
\beq \label{36}
\Lambda_h (X) \simeq  \frac{\ah}{\ln^2 X}
\eeq
where
$$
\ah = \alpha \prod_{p|h}\(1+\frac{1}{p-2}\)
\;\;\;\mbox{and }\;\;\;\alpha=2\prod_{q>2}\(1-\frac{1}{(q-1)^2}\).
$$
The first product is taken over primes which divide $h$ and the second one
is taken over all primes (except 2).

$\ah$ is an irregular number--theoretic function, and its main contribution
comes from its average behavior, $\langle \ah \rangle$. Moreover, we will only
need the average behaviour of $\ah$ for large values of $h$ (cf Eq.(\ref{311})
below). If one averages over {\em all} integers \cite{keating}
\beq \label{37}
\langle \ah \rangle \simeq 1 - \frac{1}{2h} \;\;\;\;\;\;\;\;\;\; {\rm as} \; h
\rightarrow \infty,
\eeq
expressing a repulsion between prime numbers for long distances.

Contrary to the Riemann case, to compute the two--point correlation function
for Dirichlet $L$--functions one has to compute the mean value of $\ah$ not
averaging over all integers but over all equal integers mod $k$ (i.e., all
integers having the same remainder mod $k$). This is important because the
quantity $\cp {\bar \chi} (p+h)$ which enters Eq.(\ref{35}) depends only on $h$
mod $k$. The details of this calculation are given in appendix A. The result is
quite simple: for large $h$
\beq \label{38}
\langle \ah \rangle \approx \left\{ \begin{array}{ll}
           1-1/2h    & \mbox{if $h\equiv 0$ mod $k$} \\
           1         & \mbox{otherwise.} \end{array} \right.
\eeq

This equation is the essential ingredient of our proof. It expresses a
non--trivial number--theoretic property of $\langle \ah \rangle$ and completely
eliminates the dependence on the character in (\ref{35}). This is due to the
fact that the $h$--independent (Poissonian) terms in (\ref{38}) introduce no
correlations between prime numbers. Then if in Eq.(\ref{35}) $\cp {\bar \chi}
(p+h)$ is replaced by its average, the sum over $h$ of the Poissonian
components vanishes. We thus only need to consider the terms $h\equiv 0$ mod
$k$, and for them $|\chi (p)|^2=1$ if $(p,k)=1$ and zero otherwise.

The computations are now exactly the same as for the Riemann
case~\cite{keating}. We briefly outline the main steps in appendix B, leading
to
\beq \label{39}
\rdo_{\rm off} \approx \frac{1}{2\pi^2} \frac{\cos \(2\pi d_{\rm av}
\eps\)}{\eps^2}.
\eeq
The final result for the two--point correlation function is, from (\ref{32}),
(\ref{34}) and (\ref{39})
\beq \label{315}
\rd = d^2_{\rm av}+\rdo_{\rm diag}+\rdo_{\rm off} \approx d^2_{\rm av}-
      \frac{\sin^2 \( \pi d_{\rm av} \eps\)}{\pi^2\eps^2},
\eeq
which coincides with the GUE two--point correlation function (\ref{11}), once
the average density is set to one. Eq.(\ref{315}) holds for an arbitrary
modulus and character.

In order to illustrate this result we have numerically computed, using the
approximate functional equation (\ref{212}), the zeros of Dirichlet
$L$--functions for several values of $k$ and, for each $k$, for several
characters (complex and real) and verified in all cases the agreement with the
GUE statistics. For example, in Fig.1a we show $\rd$ for the approximately
$18000$ zeros lying in the interval $10^5 \leq t \leq 1.1\times 10^5$ for $k=5$
and character $\chi_2$ of table 1. For completness, in part b of that figure we
plot the nearest--neighbour spacing distribution for those zeros; the
continuous curve is the Wigner surmise $p(s)= a s^2 \exp \( -b s^2 \)$ where
$a=32/\pi^2$ and $b=4/\pi$. Note that our result (\ref{315}) for $\rd$ does not
imply the Wigner surmise for $p(s)$, since $p(s)$ has contributions from all
$n$--point correlation functions. The agreement suggests, however, the validity
of the GUE ensemble beyond the two-point correlation function.

We shall now consider the correlations between zeros of different
$L$--functions, which are generally believed to be statistically independent.
For that purpose, we take the product of several $L$--functions having
different characters $\chi_i$ mod $k_i$. The total density is $\dav =\sum_{i}
d_{\rm av}^{i} = \sum_{i} f_{i} \dav $, where the $f_i$'s are the relative
densities and $\sum_i f_i =1$. After averaging of exponential terms, the
two--point correlation function for the product of $L$--functions can be
written as
\beq \label{316}
\rd = \dav^2 + \sum_i R_2^{{\rm osc} (i)} (\eps)
      +\sum_{i\neq j} \langle d_{\rm osc}^{(i)} (t) \, d_{\rm osc}^{(j)}
      (t+\eps) \rangle .
\eeq

Now instead of $\chi (p) \bar{\chi} (p+h)$ in Eq.(\ref{35}) we will find that
$\langle d_{\rm osc}^{(i)} (t) d_{\rm osc}^{(j)} (t+\eps) \rangle $ is
proportional to $\chi_i (p) \bar{\chi_j} (p+h)$. Because of Eq.(\ref{38}), only
the terms $h \equiv 0$ mod $k_j$ contribute and the result is therefore
proportional to $\chi_i (p) \bar{\chi}_j (p)$. When $i\neq j$ the last quantity
defines a non--principal character modulo the least common multiple of $k_i$
and $k_j$. All other terms are smooth in this scale and therefore the main
contribution comes from the {\em average} value of this function. Since the
mean value of any non--principal character is zero~\cite{bateman}
$$
\frac{1}{k_i} \sum_{p=0}^{k_i-1} \chi_i(p)=0,
$$
it follows that
$$
\langle d_{\rm osc}^{(i)} (t) \, d_{\rm osc}^{(j)} (t+\eps) \rangle = 0,
\;\;\;\;\;\;\;\;\;\;\;\;\; \forall \; i \neq j.
$$
We thus conclude that zeros of different $L$--functions (\ie, different $k$
and/or different character) are uncorrelated
\beq \label{317}
\rd = \dav^2 - \sum_i \frac{\sin^2 \( \pi f_i \dav \eps\)}{\pi^2\eps^2}.
\eeq

We have also numerically verified this prediction. Fig.2 plots the correlations
between zeros for the product of the (three) non--principal characters for
$k=5$. The fluctuations are much smaller than in Fig.1 because of the better
statistics. The continuous curve in part a) is the prediction (\ref{317}),
while the GUE result for $p(s)$ for a product of independent sets can be
found in the appendix $2$ of Ref.\cite{mehta}. More extensive numerical
computations concerning the critical zeros of Dirichlet $L$--functions can be
found in \cite{hejhal,rumely}.

\section{Epstein zeta function}
\setcounter{equation}{0}

Let us consider briefly another interesting example of zeta function,
namely Epstein's zeta function associated with a positive definite
quadratic form $Q(x,y)=ax^2+bxy+cy^2$
\beq \label{41}
Z_Q (s) = \sum_{m,n} Q (m,n) ^{-s},
\eeq
where the summation is done over all integers $m$, $n$ except $m=n=0$. $Z_Q
(s)$ is an analytic function, regular for $Re(s) >1$, satisfying a functional
equation and an approximate functional equation (see e.g. \cite{potter,pt}). In
the following, we will consider the particular case $a$, $b$ and $c$ integers.

The properties of these functions strongly depend on the value of the
discriminant $\Delta=b^2-4ac$. If the class number of quadratic forms with a
given discriminant is one, $Z_Q (s)$ can be written as a product of two
Dirichlet $L$--series, like for example
\beq \label{42}
\sum_{m,n} \frac{1}{\( m^2+n^2\)^s} = 4 \, \rs \, \ds
\eeq
where $\ds$ is the non--principal Dirichlet $L$--function for $k=4$ (cf. table
1)
$$
\ds = 1 - \frac{1}{3^s} + \frac{1}{5^s} - \frac{1}{7^s} + \cdots .
$$
This type of Epstein zeta functions obviously have an Euler product and are
assumed to satisfy the generalized Riemann hypothesis. For them, the results of
the previous section hold, and we arrive at the conclusion that the statistical
properties of their critical zeros are those of an uncorrelated superposition
of two GUE sets, Eq.(\ref{317}). Fig.3 illustrates this behaviour for the
function (\ref{42}).

When the class number is bigger than one, Epstein's functions cannot be
factorized as in (\ref{42}) and have no Euler product. Only a sum over all
classes of forms with a given discriminant is believed to have the
above--mentioned properties. One (simple) member of this non--Euler class of
functions is
\beq \label{43}
Z_Q (s) = \sum_{m,n} \frac{1}{\( m^2+5n^2\)^s}.
\eeq
For such functions with integer coefficients it is known
\cite{pt}-\cite{hejhal2} that
\begin{itemize}
\item[(i)] there is an infinite number of zeros lying on \cl;
\item[(ii)] many zeros lie off that line;
\item[(iii)] however, almost all the zeros lie on \cl or in its immediate
neighbourhood.
\end{itemize}

Statement (iii) was recently made more precise. In fact, in
Refs.~\cite{bombieri,hejhal2} it was proved that $N_c (t)/N(t) \rightarrow 1$
as $t\rightarrow \infty$, where $N(t)$ denotes the number of zeros of $Z_Q (s)$
whose imaginary part is less or equal to $t$ and $N_c (t)$ the fraction of them
lying on \cl. To prove this the generalized Riemann hypothesis for Dirichlet
$L$--functions was assumed to be valid as well as certain assumptions on the
correlations between zeros.

Eq.(\ref{43}) is an example of class number $2$ Epstein's zeta function, which
in general can be expressed as a sum of two terms $L_1 L_2 \pm L_3 L_4$, the
$L_i$ being appropriate Dirichlet $L$--functions. In \cite{bombieri,hejhal2} it
was proved that in sums of this type, in a certain range of $t$, typically one
of the two terms `dominate'. This suggests that the statistical properties of
the critical zeros for functions of the type (\ref{43}) should be close to an
uncorrelated superposition of two GUE sets.

\section{Concluding Remarks}

We have  computed the two--point correlation function for the critical zeros of
Dirichlet $L$--functions using the Hardy--Littlewood conjecture for the
distribution of prime numbers and showed that for any modulus and character the
main term agrees with the statistics of the Gaussian Unitary Ensemble of random
matrices. These results generalized those of Ref.~\cite{gm,keating} for the
Riemann zeta function, and provide a unifying property for all $L$--functions.

The problem of estimating the next--to--leading terms (which should tend to
zero as $t\rightarrow \infty$) is not simple, since it is connected to the
short--range correlations between prime numbers~\cite{montgomery,gm}.

The Hamiltonian matrix of a quantum system having a discrete symmetry (like
parity) splits, in the appropriate basis, into uncoupled submatrices, each of
them corresponding to a symmetry class. It is well known (but not proved) that
for a dynamical system with at least two degrees of freedom the eigenvalues
belonging to different submatrices are uncorrelated. To each of these
submatrices is associated a character of the symmetry group, which enters in
the semiclassical trace formula. In fact, the symmetry-projected semiclassical
spectral density includes the character of the symmetry group \cite{robbins}
exactly in the same way as in Eq.(2.7b) for Dirichlet $L$--functions. Thus,
different characters in $L$--functions are the analog of different symmetry
classes in quantum mechanics. In the light of these considerations, our result
on the statistical independence of the zeros of different $L$--functions can be
interpreted as the analog of the above-mentioned independence of the
eigenvalues of different symmetry classes in quantum mechanics.

In all the cases we have investigated the zeros of zeta functions were
distributed according to the statistics of the Gaussian {\em Unitary} Ensemble
of random matrices, or a superposition of a few of them. If there exists a
number--theoretic zeta function whose zeros obey the statistics of the
Gaussian {\em Orthogonal} Ensemble remains an open problem.

\vspace{0.4in}

\noindent\undertext{\bf Acknowledgements}: We acknowledge discussions with O.
Bohigas and M. C. Gutzwiller at the initial stage of this work, and one of the
referees for pointing out the reference \cite{rumely}. One of us (PL) thanks
the Department of Physics of the University of Buenos Aires where part of the
paper was written.

\vspace{0.4in}

\renewcommand{\appendix}
        {
        \par
        \setcounter{section}{0}
        \setcounter{subsection}{0}
        \gdef\afterthesectionpunctdefault{:}
        \gdef\thesection{{Appendix \Alph{section}}}
         \renewcommand{\theequation}{\Alph{section}.\arabic{equation}}
        \setcounter{equation}{0}
        }

\appendix
\section{}
\setcounter{equation}{0}

In this appendix we prove Eq.(\ref{38}), concerning the average behaviour of
$\ah$ mod $k$. The function $\ah$  in (\ref{36}) can be expressed
as~\cite{keating}
\beq \label{a1}
\alpha (h) = \alpha \sum_{d|h} \beta (d)
\eeq
where the sum runs over all divisor of $h$ and $\beta (d) =0$
if $d$ is even or divisible by the square of a prime, otherwise
$$
\beta (d)=\prod_{p|d}\frac{1}{p-2}.
$$
We are interested in the average over all integers mod $k$. To do this let
us define
\beq \label{a1b}
\langle \ah \rangle \equiv \frac{d}{dN} f(N)|_{N=h}
\eeq
where
$$
f(N)=\sum_{n=0}^N \alpha (nk +q).
$$
{}From (\ref{a1}) it follows that
\beq \label{a2}
f(N) = \alpha \sum_{d=1}^X {\cal N}(N,d) \beta (d)
\eeq
where $X=Nk+q$ and ${\cal N}(N,d)$ is the number of terms in the sequence
$nk+q$, $n=0,\cdots,N$ divisible by $d$.

Assuming that $(d,k)=1$ it is easy to see that
$
{\cal N}(N,d)=\[ \( N+N_o\)/d \]
$
where $N_o = \( \( 1/k\)_d q \)_d$ (henceforth, square brackets denote integer
part, curly brackets fractional part and $(x)_l$ means $x$ mod $l$).

Eq.(\ref{a2}) can be rewritten as the sum of  two terms
\beq \label{a3}
f(N) = \alpha \sum_{d=1}^X \(\frac{ N+N_o}{d }\)
\beta (d) - \alpha \sum_{d=1}^X \cbl \frac{N+N_o}{d}\cbr \beta (d).
\eeq
The term proportional to $N$ in the first sum can be computed from the relation
\cite{keating}
$$
\sum_{d=1}^{\infty} \frac{\beta (d)}{d} = \frac{1}{\alpha};
$$
then , in the large--$N$ limit
\beq \label{a4}
\alpha N \sum_{d=1}^X \frac{\beta (d)}{d} \simeq N.
\eeq
In the second sum, the fractional part depends only on $N$ mod $d$  and its
average value equals $1/2$. Moreover, because as $d\rightarrow \infty$ $\beta
(d)$ behaves, on average, like $(\alpha d)^{-1}$ (see Ref.\cite{keating}) then
\beq \label{a5}
\alpha \sum_{d} \cbl \frac{N+N_o}{d}\cbr \beta (d) \simeq \frac{\alpha}{2}
\sum_{d=1}^X \beta (d) \simeq \frac{1}{2} \log N.
\eeq

The most delicate term is the one involving $N_o$ in the first sum, since
$N_o$ is a function of $d$
$$
\alpha \sum_{d=1}^X N_o \frac{\beta (d)}{d} = \alpha \sum_{d=1}^X
\( m q \)_d \frac{\beta (d)}{d},
$$
where $m=\(1/k\)_d$. Since by definition $m k =1$ mod $d$, one can write
\beq \label{a6}
mk = 1+nd,
\eeq
$n$ being an integer ($<k$) depending only on $(d)_k$. Neglecting the
contribution of $q/k$, from (\ref{a6}) it follows that
\beq \label{a7}
\(mq\)_d \simeq \cbl \frac{n q}{k}\cbr d.
\eeq
But again, on average, we can do the replacement
$$
\cbl \frac{n q}{k}\cbr \rightarrow \frac{1}{2}.
$$
Then
$$
\alpha \sum_{d=1}^X \( \(
\frac{1}{k} \)_d q \)_d \frac{\beta (d)}{d} \simeq \alpha \sum_{d=1}^X
\cbl \frac{n q}{k} \cbr d \frac{\beta (d)}{d} \simeq \frac{\alpha}{2}
\sum_{d=1}^X
\beta (d) \simeq \frac{1}{2} \log N,
$$
which exactly cancels the term (\ref{a5}). From this, Eq.(\ref{38}) follows.

\vspace{0.4in}

\section{}
\setcounter{equation}{0}

In this appendix we outline the main steps leading to Eq.(\ref{39}).
Replacing in (\ref{35}), as for the diagonal term, the sums by integrals and
from the considerations following Eq.(\ref{38}) we can write
\beq \label{310}
\rdo_{\rm off} \approx \frac{1}{2\pi^2} \int_1^\infty \frac{d p}{p} \me^{- \mi
\eps
\ln p} \int_2^\infty dh \langle \ah \rangle \cos \(th/p\)+{\rm c.c.}.
\eeq
We now make two changes of variables, $w=(\ln p) / 2\pi d_{\rm av}$ and
$y= (2\pi)^w t^{1-w} h$. In the first one, we include the average density
in order to set the mean level spacing between zeros to one. For the second,
since $p=\exp (2\pi d_{\rm av} w)$ and from the asymptotic result
(2.7a) for the average density, we have $t/p \simeq (2\pi )^w
t^{1-w}$. Therefore
\beq \label{311}
\rdo_{\rm off} \approx \frac{d_{\rm av}}{\pi} \int_0^\infty d w \exp \(- 2\pi
\mi
d_{\rm av} \eps w\) \frac{t^{w-1}}{(2\pi)^w} \int_{2(2\pi)^w t^{1-w}}^\infty
dy \langle \alpha \( \frac{y t^{w-1}}{(2\pi)^w}\) \rangle \cos y + {\rm c.c.}.
\eeq
The lower bound of the second integral has a different asymptotic behaviour
according to whether:
\begin{itemize}
\item[i.] $w<1$; when $t \rightarrow \infty$, $t^{1-w} \rightarrow
\infty$. The integral vanishes.
\item[ii.] $w>1$. In the limit, $t^{1-w} \rightarrow 0$.
\end{itemize}

As $t\rightarrow\infty$, the argument of the averaged $\alpha$-function
tends to infinity, and we can use the asymptotic result (\ref{37}). As stated
before, the Poissonian component does not contribute, since the integral
over that term in (\ref{311}) vanishes. The second term yields
\beq \label{312}
\rdo_{\rm off} \approx -\frac{d_{\rm av}}{2\pi} \int_1^\infty d w \exp\(- 2\pi
\mi
d_{\rm av} \eps w\) \int_{2(2\pi)^w t^{1-w}}^\infty dy \cos y / y + {\rm c.c.}.
\eeq
But, for small $x$, $\int_x^\infty dy \cos y /y \simeq - C - \ln x + {\cal
O} (x^2)$, where $C$ is Euler's constant. This approximation and the
asymptotic average density allow to rewrite (\ref{312}) as
$$
\rdo_{\rm off} = \frac{d_{\rm av}}{2\pi} \int_1^\infty d w \me^{- 2\pi \mi
d_{\rm av} \eps w} 2 \pi d_{\rm av} \( 1-w \) + {\rm c.c.}
$$
or, by an evident transformation
\beeqn \label{313}
\rdo_{\rm off} &=& \int_{d_{\rm av}}^\infty d \tau \me^{- 2\pi \mi
\tau \eps} \( d_{\rm av}-\tau \) + {\rm c.c.}\\ \label{314}
    &=& \frac{1}{2\pi^2} \frac{\cos \(2\pi d_{\rm av} \eps\)}{\eps^2}.
\eeqn

\vspace{0.4in}

\pagebreak

\pagebreak

\begin{table}[p]
\begin{tabular}{|c|c|c|c|c|} \cline{2-5}
\multicolumn{1}{c|}{}
  & k=4 & \multicolumn{3}{c|}{k=5} \\  \hline\hline
n &     &$\chi_1$&$\chi_2$&$\chi_3$\\ \hline
1 &  1  &   1    &    1   &   1    \\ \hline
2 &  0  &  -1    & $\mi$  &-$\mi$  \\ \hline
3 & -1  &  -1    &-$\mi$  & $\mi$  \\ \hline
4 &  0  &   1    &   -1   &  -1    \\ \hline\hline
\end{tabular}
\caption{Non--principal characters for $k=4$ and $5$.}
\end{table}

\pagebreak

\vspace{0.4in}

\large
\begin{center}
FIGURE CAPTIONS
\end{center}
\normalsize
\baselineskip 0.3in

\begin{description}

\item{Figure 1:} a) Two-point correlation function and b) nearest-neighbor
spacing distribution for the critical zeros of Dirichlet $L$--function with
$k=5$ and character $\chi_2$ of table 1. The continuous curves are
the GUE results.

\item{Figure 2:} Same as in Figure 1 but for the critical zeros of the
product of the three non--principal characters for $k=5$.

\item{Figure 3:} nearest-neighbor spacing distribution for the critical zeros
of Epstein's zeta function (\ref{42}) lying in the interval $10^5 \leq t \leq
1.1 \times 10^5$.
\end{description}
\end{document}